\documentclass[12pt]{article}

\usepackage[cp1251]{inputenc}  
\usepackage[T2A]{fontenc}      
\usepackage[russian]{babel}    
\usepackage{amssymb}
\usepackage{amsmath}
\usepackage[dvips]{graphicx}
\begin{document}


УДК 511.9+519.147

\begin{center}
Комбинаторная оценка мощности множества, содержащего кратчайший вектор, образованного базисом решетки 

приведенной блочным методом Коркина"--~Золотарева
\end{center}

\begin{flushright}
Кузьмин О.В., quizminov@mail.ru
\end{flushright}

\begin{flushright}
Усатюк В.С., L@Lcrypto.com
\end{flushright}

 \textit{ В работе получены комбинаторные оценка мощности множества, содержащего кратчайший вектор, образованного базисом решетки приведенной блочным методом Коркина"--~Золотарева. Показано влияние плотности решеток на рост мощности этого множества. Получены верхние и нижние оценки мощности множеств, содержащих кратчайший вектор, образованных базисами экстремальных решеток, а так же решеток Гольштейна"--~Майера приведенных блочным методом Коркина"--~Золотарева.}

Ключевые слова: поиск кратчайшего вектора, мощность пространства перебора, блочный метод Коркина"--~Золотарева, экстремальные решетки, решетки Гольдштейна"--~Майера. 

\begin{center}
Bounds on the cardinality containing shortest vectors in a lattice reduced by block Korkin"--~Zolotarev method 

\end{center}

\begin{flushright}
Kuzmin O.V., quizminov@mail.ru

Usatyuk V. S., L@Lcrypto.com
\end{flushright}

\textit{ This article present a concise estimate of bound on the cardinality containing shortest vector in a lattice reduced by block Korkin"--~Zolotarev method (BKZ) for different value of the block size. Paper show how density affect to this cardinality, in form of the ratio of shortest vector size and successive minimal. Moreover we give upper estimate of cardinality for extremal(global critical) and Goldstein"--~Mayer lattices.  }


\textbf{Введение }

В работе [1], Коркиным и Золотаревым был предложен метод приведения положительной квадратичной формы (ПКФ) от $m$-переменных. На основе перехода от ПКФ к точечным решеткам и «ослабления» метода, был построен первый полиномиальный по временной сложности алгоритм приведения базиса решеток, известный как LLL"--~алгоритм. В работе [2] был использован оригинальный метод Коркина-Золотарева, на основе перехода от ПКФ к точечным решеткам, был получен метод приведения решеток по Коркину"--~Золотареву. В работе [3] был предложен метод приведения базиса, получивший название "--~блочный метод Коркина"--~Золотарева(BKZ, Block Korkin"--~Zolotarev method), в зависимости от вариации параметров ($\delta ,\beta $), образующий иерархию методов приведения решеток от LLL"--~алгоритма ($\beta =2$) до приведения по Коркину"--~Золотареву (блока состоящего из всех векторов базиса решетки, $\beta =m$). 

\textbf{Основные понятия и обозначения}

\textbf{Определение 1.} Решетка - дискретная абелева подгруппа, заданная в пространстве $R^{n} $.

Пусть базис $B=\{ b_{1} ,...,b_{n} \} $ задан линейно-независимыми векторами в $ R^{m} $. Тогда под решеткой будем понимать множество целочисленных линейных комбинаций этих векторов: 

\[ L(b_{1} ,...,b_{n} )=\{ \sum _{i=1}^{n} x_{i} b_{i} :(x_{1} ,...,x_{n} )\in Z^{n} \}, \] где $m$ и $n$, размерность и ранг решетки соответственно, $m \geq n$. 

\textbf{Определение 2.} Решетки, у которых размерность и ранг равны, называются полными решетками. 

\textbf{Определение 3. }Решетки $L_{1} $, $L_{2} $, заданные базисами $B=\{ \bar{b}_{1} ,...,\bar{b}_{n} \} $, $B'=\{ \bar{b}'_{1} ,...,\bar{b}'_{n} \} $, конгруэнтны, $L_{1} (B)\cong L_{2} (B')$, если объемы фундаментальных параллелепипедов, образованных их базисами, равны $\det (L_{1} )=\det (L_{2} )$, где $\det L=\left|\sqrt{\det (B^{T} B)} \right|$ или $\det L=\left|\det B\right|$ для полноразмерных решеток.

Множество конгруэнтных решеток может быть получено в результате умножения базиса $B$ решетки на целочисленные унимодулярные матрицы. 

\begin{center}
\includegraphics[bb=0mm 0mm 209mm 297mm, width=87.6mm, height=52.9mm, viewport=4mm 4mm 205mm 293mm]{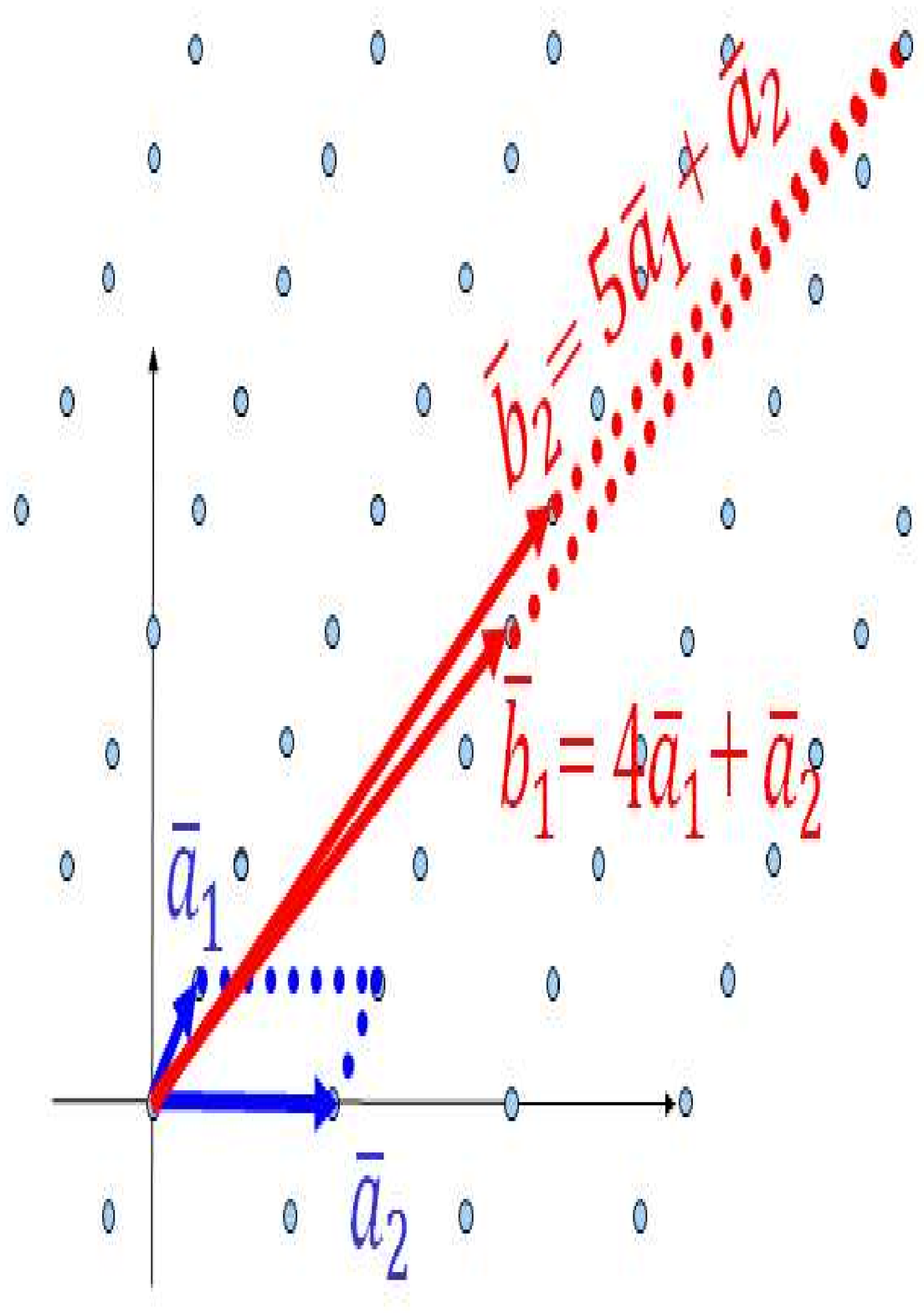}
 \end{center}
\begin{center}
 Рисунок 1. -- Конгруэнтные решетки  $L_{1} (B_{1} )=L_{2} (B_{2} )$, заданные базисами $B_{1} =\{ (1,1),(4,0)\} $ и $B_{2} =\{ (8,4),(9,5)\} $.
\end{center}
\textbf{Определение 4.} Пусть $L$ решетка ранга $n$. i-м соответствующим минимумом, где $i=1,2,...,n$,   называется точная нижняя граница радиусов замкнутых шаров с центром в нуле, содержащих $i-$линейнонезависимых векторов решетки $L$:

\[\lambda _{i} (L)=\inf \{ r|\dim (span(L\bigcap B(0,r)))\ge i\} ,\] 
где $B(0,r)=\{ x\in R^{m} |\left\| x\right\| \le r\} $ - замкнутый шар, радиуса r с центром в нуле.

\begin{center}
\includegraphics[bb=0mm 0mm 209mm 297mm, width=67.1mm, height=52.9mm, viewport=4mm 4mm 205mm 293mm]{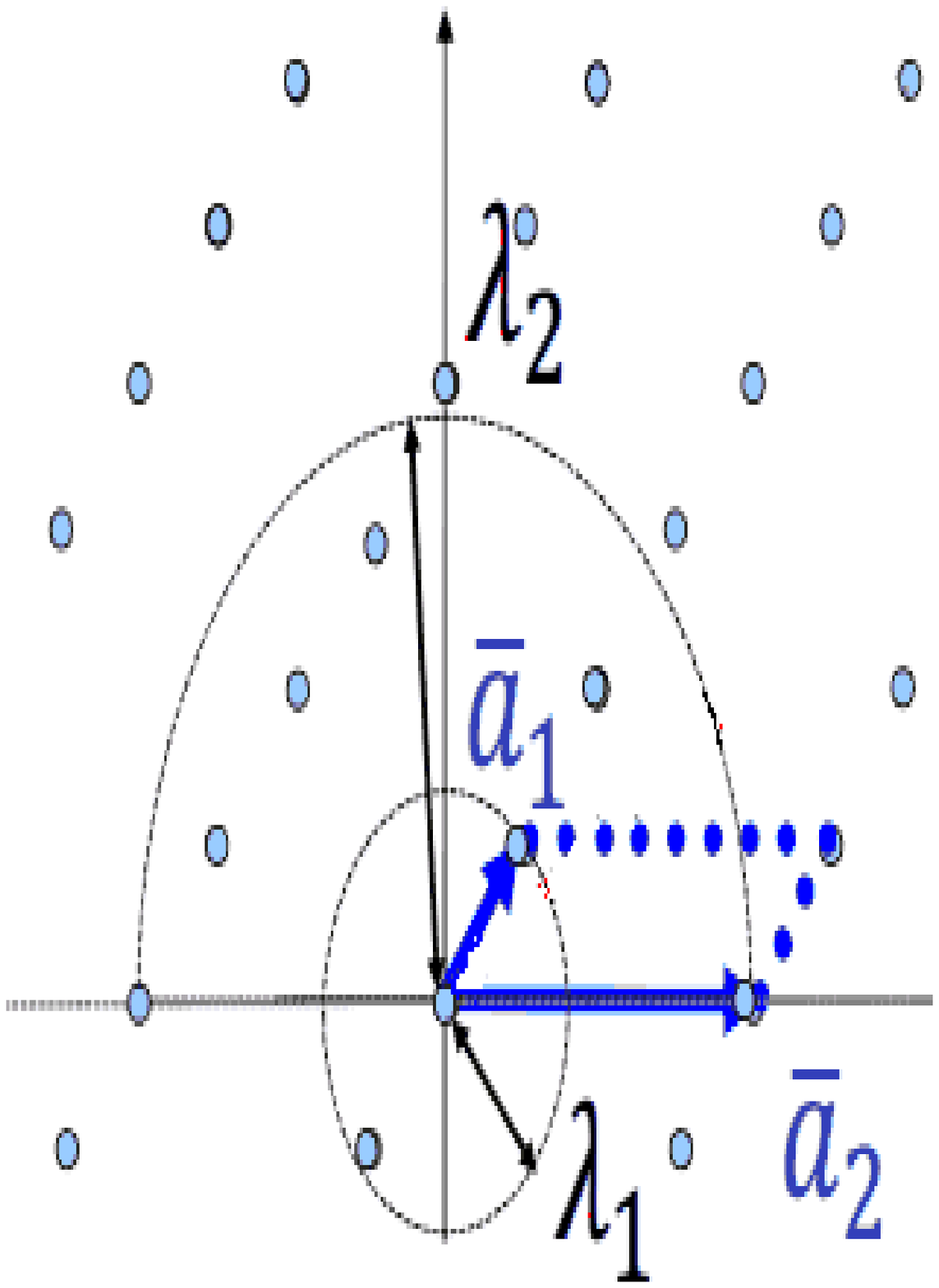}
\end{center}
\begin{center}
Рисунок 2. -- Соответствующие минимумы решетки $L_{1} (B_{1})$.
\end{center}

Длине кратчайшего вектора в решетке $L$ соответствует $\lambda _{1} (L)$, что приводит нас непосредственно к формулировке задачи поиска кратчайшего и короткого векторов в решетке.

\textbf{Определение 5.} Задача поиска короткого вектора ($\gamma $-short vector problem, ${\rm SVP} _{\gamma } (m)$): Пусть дана $m$-мерная решетка $L(B)$, ранга $n$ и вещественное $\gamma >1$. Найти нетривиальный вектор в $\gamma $-раз больший кратчайшего вектора в решетке $\bar{b}\in L :\left\| \bar{b}\right\| \le \gamma \cdot \lambda _{1} (L)$. 

При $\gamma =1$, решается задача поиска кратчайшего вектора в решетке, при $\gamma >1$"--~короткого вектора.

\textbf{Определение 6}.  Константой  Эрмита называется  величина $\gamma _{m} $, заданная выражением:

\[\gamma _{m} =\sup _{L} \left\{\frac{\lambda _{1} (L)^{2} }{(\det L)^{\frac{2}{m} } } \mid L - \text{полная} \right\}.\] 

Применяя 1 теорему о соответствующих минимумах Минковского с одной стороны и лемму о норме кратчайшего вектора в решетки, используемую в дальнейшем при доказательстве теоремы 1 получим соответственно, верхнюю  и нижнюю оценки длин кратчайшего вектора в решетке:

\[\min _{i=1}^{m} \left\| b_{i}^{\bot } \right\| \le \lambda _{1} (L)\le \sqrt{\gamma _{m} } \det (L)^{\frac{1}{m} } ,\] 
где $b_{i}^{\bot } $"--- ортогональный вектор, полученный из базиса одним из QR-методов ортогонализации [4].

\textbf{Определение 7.} Задача поиска короткого базиса решетки ($\gamma $-short basis problem, ${\rm SBP} _{\gamma } (m)$): Пусть дан базис полной решетки $B$ и вещественное $\gamma >1$. Найти базис $B'=\{ b_{1}^{'} ,b_{2}^{'} ,...,b_{m}^{'} \} :L(B)=L(B'),\prod _{i=1}^{m}\left\| b_{i}^{'} \right\|  \le \gamma \cdot \prod _{i=1}^{m}\left\| b_{i}^{\bot } \right\|  $.  

При $\gamma =1$, полученный базис достигает нижней границы неравенства Адамара и является ортогональным: 

\[\prod _{i=1}^{m}  \left\| b_{i}^{\bot } \right\| _{p} =\det \left(L\right)\le \prod _{i=1}^{m}  \left\| b_{i} \right\| _{p}. \]

\textbf{Определение 8. } Базис  $B=\{ b_{1} ,b_{2} ,\ldots ,b_{m} \} $ решётки $L\subset R^{m} $ приведён по длине, если   в результате ортогонализации решетки методом Грамма-Шмидта выполняется следующее неравенство:
\[\left|\mu _{i,j} \right|\le \frac{1}{2} ,1\le j<i\le m,\] 
где $\mu _{i,j} $- коэффициенты Грамма-Шмидта.

\textbf{Определение 9. }Упорядоченный по длине базис $B=\{ b_{1} ,b_{2} ,\ldots ,b_{m} \} $ решётки $L\subset R^{m} $ приведён блочным методом Коркина"--~Золотарева  с блоком $\beta \in \left[2,m\right]$ и точностью $\delta \in \left(\frac{1}{2} ;1\right]$, если:   

базис $B$ приведён по длине;  

$\delta ^{2} \cdot \left\| b_{i}^{\bot } \right\| ^{2} \le \lambda _{1}^{2} (L_{i} ),i=1,\ldots ,m,$ где $\lambda _{1} (L_{i} )$-длина кратчайшего вектора в решётке $L_{i} $, образованной ортогональным дополнением векторного пространства с базисом $b_{i} ,\ldots ,b_{\min (i+\beta -1,m)} $.  

\textbf{Определение 10. }Базис решетки приведен по Ленстра"--~Ленстра"--~Ловасу (LLL--алгоритмом), если он приведен блочным методом Коркина-Золотарева с размером блока $\beta =2$ и точностью $\delta \in \left(\frac{1}{2} ;1\right)$.

Отметим, что в классической работе [8], $\delta ^{2} =\frac{3}{4} $. На практике можно взять $\delta ^{2} =1$, однако LLL"---алгоритм приведения базиса перестает быть полиномиальным по временной сложности.\textbf{}

\textbf{Определение 11. }Семейство $\ell =\{ C_{1} ,C_{2} ,\ldots \} $ компактных множеств с непустой внутренностью называется упаковкой в $R^{n} $, если

\[\Omega =\mathop{\bigcup }\limits_{i} C_{i} \subseteq R^{n} .\] 

и никакие два из множеств $C_{i} $ не имеют общей внутренней точки.

\textbf{Определение 12. }Если упаковка $\ell $ в $R^{n} $ состоит из копий некоторого измеримого  множества $C\subset R^{n} $, расположенных в каждой точке решетки $L$, то есть  $\ell =\{ C+a|a\in L\} $, то $\ell $(а также множество $\Omega $) называется решетчатой упаковкой. 

\textbf{Определение 13. }Под плотностью решетки $L$ понимают величину равную отношению объема  шара радиуса равного половине кратчайшего вектора решетки к объему этой решетки: 

\[\Delta (L)=(\det L)^{-1} \cdot V_{m} (\frac{\lambda _{1}(L) }{2} ).\] 
Минковский в своей неопубликованной работе 1905 г., неконструктивно доказал, что существуют решетки с плотностью $\Delta (L)\ge 2^{-m+1} $. В теореме Минковского"--~Главки результат был уточнен $\Delta (L)\ge \frac{\zeta (m)}{2^{m-1} } $, где $\zeta (m)$- дзета-функция Римана, при $m\to \infty $ стремящейся к единице [5]. 
\textbf{Определение 14. }Решетчатая упаковка $\ell $ является плотнейшей (наиплотнейшей), если

\[\Delta (L)=\mathop{\sup } \left\{(\det L)^{-1} \cdot V_{m} (\frac{\lambda _{1} }{2} )|\forall L:\dim (L)=m\right\}.\] 

Иными словами плотнейшей (наиплотнейшей) решетчатой упаковке соответствует, такое размещение шаров с центрами в точках решетки, которое позволяет покрыть наибольшую часть фундаментального парраллелепипеда образованного этой решеткой, среди всевозможных решеток которые могут быть построенны в данной размерности, (рис.3).

Эта задача является одной из классических задач комбинаторной геометрии. 

\begin{center}
\includegraphics[bb=0mm 0mm 209mm 297mm, width=58.4mm, height=52.5mm, viewport=4mm 4mm 205mm 293mm]{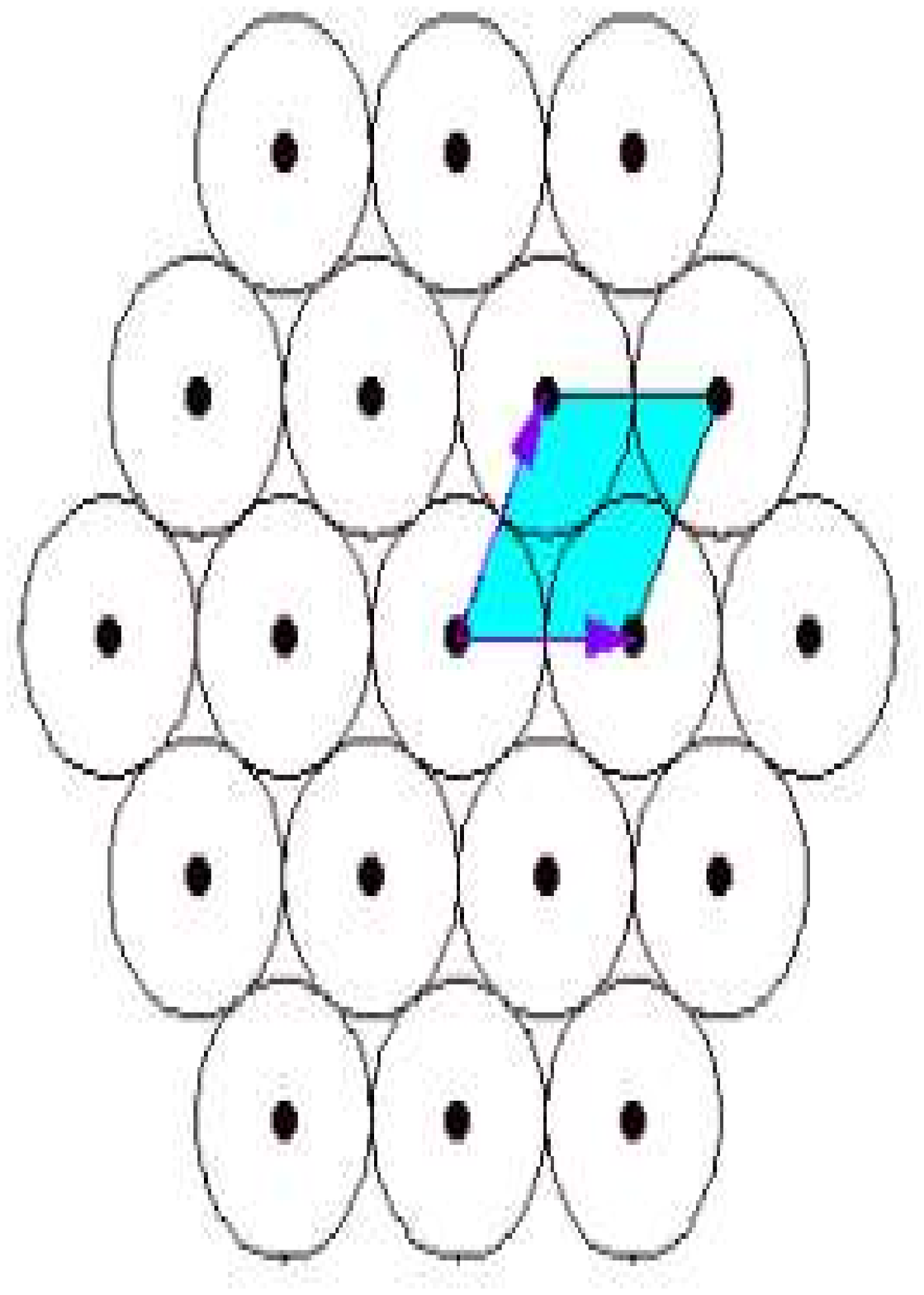}
\end{center}

\begin{center}
Рисунок 3. -- Решетчатая упаковка шаров в гексагональной решетке $L(B)$ образованной базисом $B=\{ (1,0),(0.5,\frac{\sqrt{3} }{2} )\} $.
\end{center}

\textbf{Определение 15. }Решетка для которой выполняется $\lambda _{1} (L)=\sqrt{\gamma _{m} }\times  \det (L)^{\frac{1}{m} } $  называется глобально \textit{критической} или \textit{экстремальной} [5,6]. 

Заметим, что константа Эрмита достигает верхней границы теоремы Минковского о соответсвующих минимума.

В соответствии с 2 теоремой о соответствующих минимумах Минковского $\prod _{i=1}^{m} \lambda _{i} (L)\le \sqrt{\gamma ^{m} _{m} } \det (L)$ легко убедиться, что все соответствующие минимумы экстремальной решетки равны друг другу $\lambda _{1} =\lambda _{2} =...=\lambda _{m} $ и ее упаковка  шарами радиуса $\frac{1}{2} \lambda _{1} $ является плотнейшей.  В таблице 1 приведены плотности и значения константы Эрмита для известных экстремальных решеток [5].

\begin{flushright}
Таблица 1
\end{flushright}
\begin{center}
  Параметры наиплотнейших решетчатых упаковок
\end{center}
\begin{center}
\begin{tabular}{|p{1.05in}|p{0.4in}|p{0.3in}|p{0.3in}|p{0.3in}|p{0.4in}|p{0.4in}|p{0.4in}|p{0.4in}|} \hline 
Размерность решетки, m & 2 & 3 & 4 & 5 & 6 & 7 & 8 & 24 \\ \hline 
Плотность  решетки, $\Delta (L)$ & 0.9069 & 0.74 & 0.62 & 0.46 & 0.373 & 0.295 & 0.254 & 0.0019 \\ \hline 
Константа Эрмита, $\gamma _{m} $ & $\frac{2}{\sqrt{3} } $ & $\sqrt[{3}]{2} $ & $\sqrt[{4}]{4} $ & $\sqrt[{5}]{8} $ & $\frac{2}{\sqrt[{6}]{3} } $ & $\sqrt[{7}]{64} $ & 2 & 4 \\ \hline 
\end{tabular}
\end{center}

\textbf{Определение 16. }Целочисленная решетка $L(B)$, чья базисная матрица задана Эрмитовой нормальной формой вида:

\[B=\left(\begin{array}{ccccc} {n} & {x_{1} } & {\cdots } & {x_{m-1} } & {x_{m} } \\ {0} & {1} & {\cdots } & {0} & {0} \\ {\vdots } & {\vdots } & {\ddots } & {\vdots } & {\vdots } \\ {0} & {0} & {\cdots } & {1} & {0} \\ {0} & {0} & {\cdots } & {0} & {1} \end{array}\right),\] 
где $n-$простое число, $x_{i}$ выбрано из равномерного распределения заданного на отрезке $[0,n-1]$, называется решеткой Гольштейна"--~Майера, [7].

Решетки Гольдштейна"--~Майера позволяют строить случайные  целочисленные решетки c плотностью «близкой» к экстремальным решеткам [8] и обладают следующим  свойством: \[\lambda _{i} (L(B))\approx \sqrt{\gamma _{m} } \det L^{\frac{1}{m} } =\sqrt{\frac{1}{\pi } \Gamma \left(1+\frac{m}{2} \right)^{\frac{2}{m} } } \det L^{\frac{1}{m} } =\sqrt{\frac{m}{2\pi e} } \det L^{\frac{1}{m} } +o(m)\]

\textbf{Оценки}

 Оценим мощность пространства перебора для решения задачи поиска кратчайшего вектора решетки, заданного базисом, предварительно приведенным BKZ"---методом.

\textbf{Теорема 1.} Пусть $b_{1} ,b_{2} ,...,b_{m} $- базис полной решетки $L\subseteq R^{m} $, который приведен BKZ"---методом с размером блока $\beta $ и $\delta =1$, $\lambda _{i} -$ i-й соответствующий минимум решетки, $\gamma _{\beta } $- константа Эрмита, $\left[\cdot \right]$-целая часть числа. Нижняя оценка мощности множества содержащего кратчайший вектор решетки $L$ равна:

\[\prod _{j=1}^{m} \left(1+2\left[\gamma _{\beta }^{\frac{j-m}{\beta -1} } \frac{\lambda _{1} \left(L\right)}{\lambda _{j} \left(L\right)} \right]\right).\] 

\textbf{Доказательство.} Рассмотрим случай полной решетки. Возьмем произвольный вектор $x\in L$, по определению решетки он представим в виде $x=\sum _{i=1}^{m} y_{i} b_{i} ,y_{i} \in Z$.

Пусть $j\le m$- наибольшее такое, что $y_{j} \ne 0$. Тогда рассмотрим модуль скалярного произведения $\sum _{i=1}^{m} y_{i} b_{i} $ на $b_{j}^{\bot } $, где $b_{j}^{\bot } $-ортогональный вектор полученный в результате применения QR-разложения, например методом ортогонализации Грама"--~Шмидта.

Поскольку скалярное произведение ортогональных векторов будут давать 0, ортогональное преобразование сохраняет скалярное произведение векторов $\left\langle b_{j} ,b_{j}^{\bot } \right\rangle =\left\langle b_{j}^{\bot } ,b_{j}^{\bot } \right\rangle $) получаем:

\[\left|\left\langle \sum _{i=1}^{j} y_{i} b_{i} ,b_{j}^{\bot } \right\rangle \right|=\left|y_{j} \right|\left\langle b_{j}^{\bot } ,b_{j}^{\bot } \right\rangle =\left|y_{j} \right|\left\| b_{j}^{\bot } \right\| ^{2} .(1)\] 

С другой стороны применим неравенство Коши"--~Буняковского получим:

\[\left|\left\langle \sum _{i=1}^{m} y_{i} b_{i} ,b_{j}^{\bot } \right\rangle \right|\le \left\| \sum _{i=1}^{m} y_{i} b_{i} \right\| \cdot \left\| b_{j}^{\bot } \right\| .(2)\] 

Подставив соотношение (1) в неравенство (2), получим:

\[{\rm \parallel }x{\rm \parallel }=\left\| \sum _{i=1}^{m} y_{i} b_{i} \right\| \ge \frac{\left|\left|y_{j} \right|\left\| b_{j}^{\bot } \right\| ^{2} \right|}{\left\| b_{j}^{\bot } \right\| } \ge \left|y_{j} \right|\left\| b_{j}^{\bot } \right\| .\]

 Последнее неравенство устанавливает нижнюю оценку длины кратчайшего вектора в решетке $\left\| x\right\| \ge \left|y_{j} \right|\left\| b_{j}^{\bot } \right\| ,y_{j} \in Z\backslash \{ 0\} $.

Аналогичным образом рассмотрим случай $y_{j-1} $. 

\[\left|\left\langle \sum _{i=1}^{j} y_{i} b_{i} ,b_{j-1}^{\bot } \right\rangle \right|=\left|\left|y_{j-1} \right|\left\langle b_{j-1}^{\bot } ,b_{j-1}^{\bot } \right\rangle +\sum _{i=j}^{m} \left\langle y_{i} b_{i} ,b_{j-1}^{\bot } \right\rangle \right|=\] 

\[=\left|\left|y_{j-1} \right|\left\| b_{j-1}^{\bot } \right\| ^{2} +\sum _{i=j}^{m} \left\langle y_{i} b_{i} ,b_{j-1}^{\bot } \right\rangle \right|.(3)\] 

С другой стороны применим неравенство Коши"--~Буняковского получим:

\[\left|\left\langle \sum _{i=1}^{m} y_{i} b_{i} ,b_{j-1}^{\bot } \right\rangle \right|\le \left\| \sum _{i=1}^{m} y_{i} b_{i} \right\| \cdot \left\| b_{j-1}^{\bot } \right\| .(4)\] 

Подставив соотношение (3) в неравенство (4), получим:

\[\left\| x\right\| =\left\| \sum _{i=1}^{m} y_{i} b_{i} \right\| \ge \frac{\left|\left|y_{j-1} \right|\left\| b_{j-1}^{\bot } \right\| ^{2} +\sum _{i=j}^{m} \left\langle y_{i} b_{i} ,b_{j-1}^{\bot } \right\rangle \right|}{\left\| b_{j-1}^{\bot } \right\| } \]

\[\frac{\left\| x\right\| }{\left\| b_{j-1}^{\bot } \right\| } \ge \left|\left|y_{j-1} \right|+\frac{\sum _{i=j}^{m} \left\langle y_{i} b_{i} ,b_{j-1}^{\bot } \right\rangle }{\left\| b_{j-1}^{\bot } \right\| ^{2} } \right|\]

\[\frac{\left\| x\right\| }{\left\| b_{j-1}^{\bot } \right\| } \ge \left|\left|y_{j-1} \right|+\sum _{i=j}^{m} y_{i} \mu _{i,j} \right|,\]

 где $\mu _{i,j} =\frac{\left\langle b_{i} ,b_{j}^{\bot } \right\rangle }{\left\langle b_{j}^{\bot } ,b_{j}^{\bot } \right\rangle } =\frac{\left\langle b_{i} ,b_{j}^{\bot } \right\rangle }{\left\| b_{j}^{\bot } \right\| ^{2} } ,1\le j<i\le m$ -коэффициенты Грама-Шмидта.

Тогда $\left|y_{j-1} \right|\in Z$ принимает целые значения на отрезке:

\[\sum _{i=j}^{m} y_{i} \mu _{i,j} -\frac{\left\| x\right\| }{\left\| b_{j-1}^{\bot } \right\| } \le \left|y_{j-1} \right|\le \sum _{i=j}^{m} y_{i} \mu _{i,j} +\frac{\left\| x\right\| }{\left\| b_{j-1}^{\bot } \right\| } .\] 

Таким образом, в худшем случае, когда $\sum _{i=j}^{m} y_{i} \mu _{i,j} -\frac{\left\| x\right\| }{\left\| b_{j-1}^{\bot } \right\| } \ge 0$, $y_{j-1} $ принимает $\frac{\left\| x\right\| }{\left\| b_{j-1}^{\bot } \right\| } $ целых значений от точки центра с координатами $\sum _{i=j}^{m} y_{i} \mu _{i,j} $.

Рассмотрим произвольный целочисленный вектор $y=\left(\begin{array}{c} {y_{1} } \\ {y_{2} } \\ {\vdots } \\ {y_{m} } \end{array}\right)$, умножение которого на векторы базиса решетки $L$, даст нам всевозможные векторы, принадлежащие этой решеткe, $x=\sum _{i=1}^{m} y_{i} b_{i} ,y_{i} \in Z$. Рассматривая координатные компоненты от m до 1 получим $\left(1+2\left[\frac{\left\| x\right\| }{\left\| b_{m}^{\bot } \right\| } \right]\right)\cdot \left(1+2\left[\frac{\left\| x\right\| }{\left\| b_{m-1}^{\bot } \right\| } \right]\right)\cdot \ldots \cdot \left(1+2\left[\frac{\left\| x\right\| }{\left\| b_{1}^{\bot } \right\| } \right]\right)=\prod _{j=1}^{m} \left(1+2\left[\frac{\left\| x\right\| }{\left\| b_{j}^{\bot } \right\| } \right]\right)$ вариантов для вектора $y$.

Так как базис фиксирован $B=\{ b_{1} ,b_{2} ,...,b_{m} \} $, то число возможных вариантов значений принимаемых векторами решетки $x=\sum _{i=1}^{m} y_{i} b_{i} ,y_{i} \in Z$ определятся мощностью целых векторов $y$ и равно 

\[\prod _{j=1}^{m} \left(1+2\left[\frac{\left\| x\right\| }{\left\| b_{j}^{\bot } \right\| } \right]\right).(5)\] 

В базисе приведенном BKZ-методом первый вектор кратчайший в блоке приведенного базиса $\left\| x\right\| \le \left\| b_{1} \right\| $, подставив соотношение в неравенство (3) получим: 

\[\prod _{j=1}^{m} \left(1+2\frac{\left\| x\right\| }{\left\| b_{j}^{\bot } \right\| } \right)\le \prod _{j=1}^{m} \left(1+2\left[\frac{\left\| b_{1} \right\| }{\left\| b_{j}^{\bot } \right\| } \right]\right).(6)\] 

Из тривиальных соображений: 

\[\left\| b_{1} \right\| \ge \lambda _{1} (L).(7)\] 

В работе [9] доказано, что в результате применения блочного метода Коркина-Золотарева для решетки $L\in R^{n} $ ранга $m$, с размером блока $\beta $, будет получен базис, $B=\{ b_{1} ,b_{2} ,...,b_{m} \} $ для которого выполняются следующие неравенства ($i=1,...,m$):

\[\left\| b_{i}^{\bot } \right\| ^{2} \le \gamma _{\beta }^{2\frac{m-i}{\beta -1} } \cdot \lambda _{i} (L)^{2} .(8)\] 

Подставив в неравенство (6), выражения (7) и (8), получим искомую оценку 

\[\prod _{j=1}^{m} \left(1+2\left[\gamma _{\beta }^{\frac{j-m}{\beta -1} } \frac{\lambda _{1} \left(L\right)}{\lambda _{j} \left(L\right)} \right]\right)\] 

\textbf{Теорема 2.} Пусть $b_{1} ,b_{2} ,...,b_{m} $- базис полной экстремальной решетки или решетки Гольштейна-Майера $L\subseteq R^{m} $, который приведен BKZ-методом с размером блока $\beta $ и $\delta =1$, $\gamma _{\beta } $- константа Эрмита, $\left[\cdot \right]$-округление до целого значения. Нижняя оценка мощности множества содержащего кратчайший вектор решетки $L$ равна:

\[\prod _{j=1}^{m} \left(1+2\left[\gamma _{\beta }^{\frac{j-m}{\beta -1} } \right]\right).\] 

\textbf{Доказательство.} Из теоремы 1 следует, что мощность пространства перебора для нахождения кратчайшего вектора, уменьшается при $\lambda _{1} (L)<\lambda _{j} (L)$, так как, в этом случае $\frac{\lambda _{1} (L)}{\lambda _{j} (L)} <1$. Для экстремальных решеток, а так же «близких» к ним, решеток сложных по Гольштейну-Майеру, выполняется $\frac{\lambda _{1} (L)}{\lambda _{j} (L)} =1$ и $\frac{\lambda _{1} (L)}{\lambda _{j} (L)} \approx 1$ соответственно. Из теоремы 1 получим мощность множества $\prod _{j=1}^{m} \left(1+2\left[\gamma _{\beta }^{\frac{j-1}{\beta -1} } \right]\right)$ векторов.\\ \\ \\

\textbf{Теорема 3.} Пусть $b_{1} ,b_{2} ,...,b_{m} $- базис полной решетки $L\subseteq R^{m} $, который приведен BKZ"---методом с размером блока $\beta $ и $\delta =1$, $\lambda _{i} -$ i-й соответствующий минимум решетки, $\gamma _{\beta } $- константа Эрмита, $\left[\cdot \right]$-целая часть числа. Верхняя оценка мощности множества содержащего кратчайший вектор решетки $L$ равна:

\[\prod _{j=1}^{m}\left(1+2\left[\gamma _{\beta }^{\frac{m+j-2}{\beta -1} } \frac{\lambda _{1} (L)}{\lambda _{j} (L)} \right]\right). \] 

\textbf{Доказательство.}
Выполняется по аналогии с доказательством теоремы 1, до получения неравенства (6):

\[\prod _{j=1}^{m} \left(1+2\frac{\left\| x\right\| }{\left\| b_{j}^{\bot } \right\| } \right)\le \prod _{j=1}^{m} \left(1+2\left[\frac{\left\| b_{1} \right\| }{\left\| b_{j}^{\bot } \right\| } \right]\right).\]

Оценим отношение $ \frac{\left\| b_{1} \right\| ^{2} }{\left\| b_{i}^{\bot } \right\| ^{2} } $.
В работе [9] доказано, что в результате применения блочного метода Коркина-Золотарева для решетки $L\in R^{n} $ ранга $m$, с размером блока $\beta $, будет получен базис, $B=\{ b_{1} ,b_{2} ,...,b_{m} \} $ для которого выполняются следующие неравенства ($i=1,...,m$):
\[\left\| b_{1} \right\| ^{2} \lambda _{1}^{-2} \le \gamma _{\beta }^{2\frac{m-1}{\beta -1} } , (9)\] 

\[\left\| b_{i}^{\bot } \right\| ^{2} \lambda _{1}^{-2} \ge \gamma _{\beta }^{-2\frac{i-1}{\beta -1} } . (10)\] 

Поделив неравенства (9) и (10) и подставив результат в выражение (6) получим искомую оценку.

\textbf{Теорема 4.} Пусть $b_{1} ,b_{2} ,...,b_{m} $- базис полной экстремальной решетки или решетки Гольштейна-Майера $L\subseteq R^{m} $, который приведен BKZ-методом с размером блока $\beta $ и $\delta =1$, $\gamma _{\beta } $- константа Эрмита, $\left[\cdot \right]$-округление до целого значения. Верхяя оценка мощности множества содержащего кратчайший вектор решетки $L$ равна:

\[\prod _{j=1}^{m}\left(1+2\left[\gamma _{\beta }^{\frac{m+j-2}{\beta -1} } \right]\right) \] 

\textbf{Доказательство.}
Выполняется по аналогии с доказательством теоремы 2.\\

\textbf{Приложение полученных результатов}

Продемонстрируем преимущество полученной оценки. При размере блока $\beta =2$, BKZ"---метод переходит в LLL"---алгоритм(Ленстра"--~Ленстра"--~Ловаса), при $\delta =1$. Однако, в этом случае LLL"---алгоритм не всегда выполняется за полиномиальное время, а значит, утрачивает свое основное преимущество. Рассмотрим классический вариант LLL-алгоритма, в котором $\delta =\frac{3}{4} $ [10]. Для него выполняется: $b_{1} =b_{1}^{\bot } $, $\left\| b_{1} \right\| \le \sqrt{2} ^{i-1} \left\| b_{i}^{\bot } \right\| ,i=2...m.$ Из доказательства теоремы 1 получим ${\kern 1pt} \prod _{i=1}^{m} \left(1+2\left[\frac{\left\| b_{1} \right\| }{\left\| b_{i}^{\bot } \right\| } \right]\right)\le 3\cdot \prod _{i=2}^{m} \left(1+2\cdot \left[\sqrt{2} ^{i-1} \right]\right)$, что лучше $(2\cdot 2^{\frac{m(m-1)}{4} } +1)^{m} $, оценки приведенной в работе (см. [11], c. 370-371).

При $\beta =m$, BKZ-метод переходит в приведение базиса по Коркину-Золотареву, [2]. Из доказательства теоремы 1 и оценок $\left\| b_{i}^{\bot } \right\| \ge \lambda _{1} (L)\cdot i^{-(1+\log i)} $ (см. [2]), $\left\| b_{1} \right\| \le \gamma _{m} \cdot \lambda _{1} (L)$ (см. [9]), получим верхнюю оценку мощности множества содержащего кратчайший вектор равную: \[\prod _{j=1}^{m} \left(1+2\left[\gamma _{m} \cdot j^{1+\log j} \right]\right). \]

Оценки получили экспериментальное подтверждение на международных конкурсах алгоритмов поиска кратчайшего вектора на  решетках Гольдштейна-Майера [12] и решетках построенных на идеале кругового многочлена [13] с использование разработанного авторами программного комплекса для решения задач приведения базиса и поиска кратчайшего вектора в решетке [14, 15].\\

\textbf{Заключение}

Доказанная обобщенная оценка демонстрирует влияние плотности решетки на рост мощности пространства перебора при решении задачи поиска кратчайшего вектора в случае предварительного приведения ее базиса блочным методом Коркина-Золотарева. Может применяться для случайных решеток, если удается вычислить отношения ее соответствующих минимумов.

\textbf{}

\textbf{Литература}

1. Korkine A., Zolotareff  G. Sur les formes quadratuques // Math. Ann., 1873.  6. P. 366--389. 

2. Lagarias J. C., Lenstra  H. W. JR., Schnorr C. P. Korkin-Zolotarev bases and successive minima of a lattice and its reciprocal lattice //Combinatorica. 1990. V.10.  4. P. 333--348. 

3. Schnorr C. P., Euchner  M. Lattice Basis Reduction: Improved Practical Algorithms and Solving Subset Sum Problems // LNCS. 1991. V. 591. P. 68--85. 

4. Press W. H., Teukolsky S. A., Vetterling W. T. Numerical Recipes: The Art of Scientific Computing. New York: Cambridge University Press, 2007. 1262 P.

5. Conway J. H., Sloane N. J.A.  Sphere Packings, Lattices and Groups. 3rd ed. 1999 edition. Springer. 706 P.

6. Малышев А.В. Основные понятия и теоремы геометрии чисел // Труды Петрозав. Гос. ун-та. Серия «Математика». 1992.  4. C. 2--42. 

7. Goldstein D., Mayer  A. On the equidistribution of Hecke points // Forum Mathematicum. 2003. V. 15.  2. P. 165--189.

8. Nguyen P.Q., Stehle D. LLL on the average, ANTS VII, 2006, LNCS 4076, P. 238-256.

9.Schnorr C. P. Block reduced lattice bases and successive minima // Combinatorics, Probability and Computing. 1994. V. 3. P. 507--522.

10. Lenstra  A., Lenstra  H., Lovasz  L. Factoring polynomials with rational coefficients // Math. Ann., 1982. V. 261.  4. P. 515--534. 

11. Глухов М. М.,Круглов И. А., Пичкур А. Б.,Черемушкин А. В.  Введение в теоретико-числовые методы криптографии. СПб.: «Лань», 2011. 400 C.

12. http://www.latticechallenge.org/svp-challenge/index.php - Goldstein-Mayer lattice challenge (SVP). 2013.

13. http://www.latticechallenge.org/ideallattice-challenge/index.php - Ideal lattice challenge (SVP, Approx-SVP). 2013.

14. Кузьмин О.В., Усатюк В.С. Программный комплекс приведения базиса целочисленных решеток// Программные продукты и системы. 2012. №4(100). C. 180-183.

15. Усатюк В.С. Реализация параллельного алгоритма поиска кратчайшего вектора в блочном методе Коркина-Золотарева //Прикладная дискретная математика. Приложение.  2013. №6. C. 130-131.

\end{document}